\documentclass[aps,prl,reprint,two column, superscriptaddress]{revtex4-1}
\usepackage{braket}
\usepackage{leftidx}
\usepackage{mathrsfs,amsmath,amssymb}
\mathchardef\mhyphen="2D 

\usepackage{url}
\usepackage{graphicx}
\usepackage{graphics,kantlipsum}
\usepackage{stackengine}
\usepackage{color}
\usepackage{float}
\usepackage{soul}
\usepackage{kbordermatrix}

\begin{document}


\title{Three-Photon Discrete-Energy-Entangled W State in Optical Fiber}


\author{B. Fang}
\email[Corresponding author: ]{bin.fang@austin.utexas.edu}
\altaffiliation[Present address: ]{The Center for Dynamics and Control of Materials:\,\,an NSF MRSEC, The University of Texas at Austin, TX 78712, USA }
\affiliation{Department of Physics, University of Illinois at Urbana-Champaign, Urbana, IL 61801, USA}
\author{M. Menotti}
\affiliation{Department of Physics, University of Pavia, Via Bassi 6, 1-27100 Pavia, Italy}
\author{M. Liscidini}
\affiliation{Department of Physics, University of Pavia, Via Bassi 6, 1-27100 Pavia, Italy}
\author{J. E. Sipe}
\affiliation{Department of Physics, University of Toronto, 60 St. George Street, Toronto, Ontario M5S 1A7, Canada}
\author{V. O. Lorenz}
\affiliation{Department of Physics, University of Illinois at Urbana-Champaign, Urbana, IL 61801, USA}



\date{\today}

\begin{abstract}
We experimentally demonstrate the generation of a three-photon discrete-energy-entangled W state using multi-photon-pair generation by spontaneous four-wave mixing in an optical fiber. We show that by making use of prior information on the photon source we can verify the state produced by this source without resorting to frequency conversion.

\end{abstract}

\pacs{11}

\maketitle


Multiphoton entangled states are a rich resource for fundamental tests of quantum mechanics \cite{Mermin1990, Pan2000, Collins2002, Cabello2002} and enable an array of powerful quantum communication protocols \cite{Pan2012, Schwaiger2015, Malik2016}. As the maximally entangled state of one of the two classes of genuine tripartite entanglement, 
the tripartite W state has a form:
\begin{equation}
\ket{W}=\frac{1}{\sqrt{3}}(\ket{bba}+\ket{bab}+\ket{abb}),
\end{equation}
where $\ket{a}$ and $\ket{b}$ are orthogonal states. Multipartite entanglement is not a trivial generalization of bipartite entanglement \cite{Horodecki2009}, which has been extensively studied and demonstrated, and a W state cannot be transformed into a Greenberger-Horne-Zeilinger (GHZ) state \cite{Greenberger1990, PhysRevLett.82.1345} through local operations and classical communication \cite{PhysRevA.62.062314}. Compared to a three-photon GHZ state, the entanglement of a three-photon W state is more robust against loss, as the remaining two-photon system still retains some entanglement \cite{Barnea2015,Sohbi2015}. W states with large numbers of photons enable clear violations of local realism and are more distillable than GHZ states in a noisy channel \cite{SenDe:PhysicalReviewA:2003}. Although less studied than the GHZ state, the W state has been shown to have promising applications in quantum teleportation, superdense coding, and quantum key distribution \cite{Murao1999, Shi2002, Joo2003, Gorbachev2003, Agrawal:PhysicalReviewA:2006}.

To date, demonstrations of photonic W states have used the polarization degree of freedom (DOF) of photon pairs generated by spontaneous parametric down-conversion \cite{Yamamoto2002, Kiesel2003, Eibl2004, Mikami2004, PhysRevLett.95.150404}. Entanglement in polarization is not always optimal for long-distance communication through optical fibers, due to polarization mode dispersion \cite{PhysRevLett.106.080404}; it was recently proposed \cite{Menotti:2016} that a W state could be generated using the energy DOF, a possibility which has not yet been implemented. As the energy of a photon cannot be easily altered without a strong nonlinearity or modulating field, the propagation of an energy-entangled W state is inherently robust against decoherence. After transmission, if matrix transformations are required for further quantum processing, a wide range of demonstrated single-photon frequency conversion techniques \cite{PhysRevLett.68.2153, PhysRevLett.109.147404, PhysRevLett.105.093604, Agha:12}
 can be applied. Thus, it makes sense to study energy as a valuable DOF, even in the context of multipartite states.
The scheme for generating a discrete-energy-entangled W state can be based on multi-pair generation in a single nonlinear material, which makes it relatively easy to implement in bulk optics, while also potentially realizable in an integrated optics platform for enhanced scalability and efficiency \cite{Menotti:2016,Grassani:15}.

In this work, we demonstrate the generation of a three-photon discrete-energy-entangled W state in an optical fiber. Simultaneous generation of two photon pairs through spontaneous four-wave mixing (SFWM) is followed by the energy-resolved detection of one of the photons. This determines the W state that characterizes the other three photons, when post-selected on a three-fold coincidence. We verify the generation of the W state via a reduced density matrix technique \cite{Xin2017}. Making use of prior information on the photon source, and experiments that mix the channel outputs, we show that a sequence of coincidence measurements is sufficient to determine its density matrix without the need for frequency conversion.

\begin{figure}[h]
\includegraphics[width=8cm]{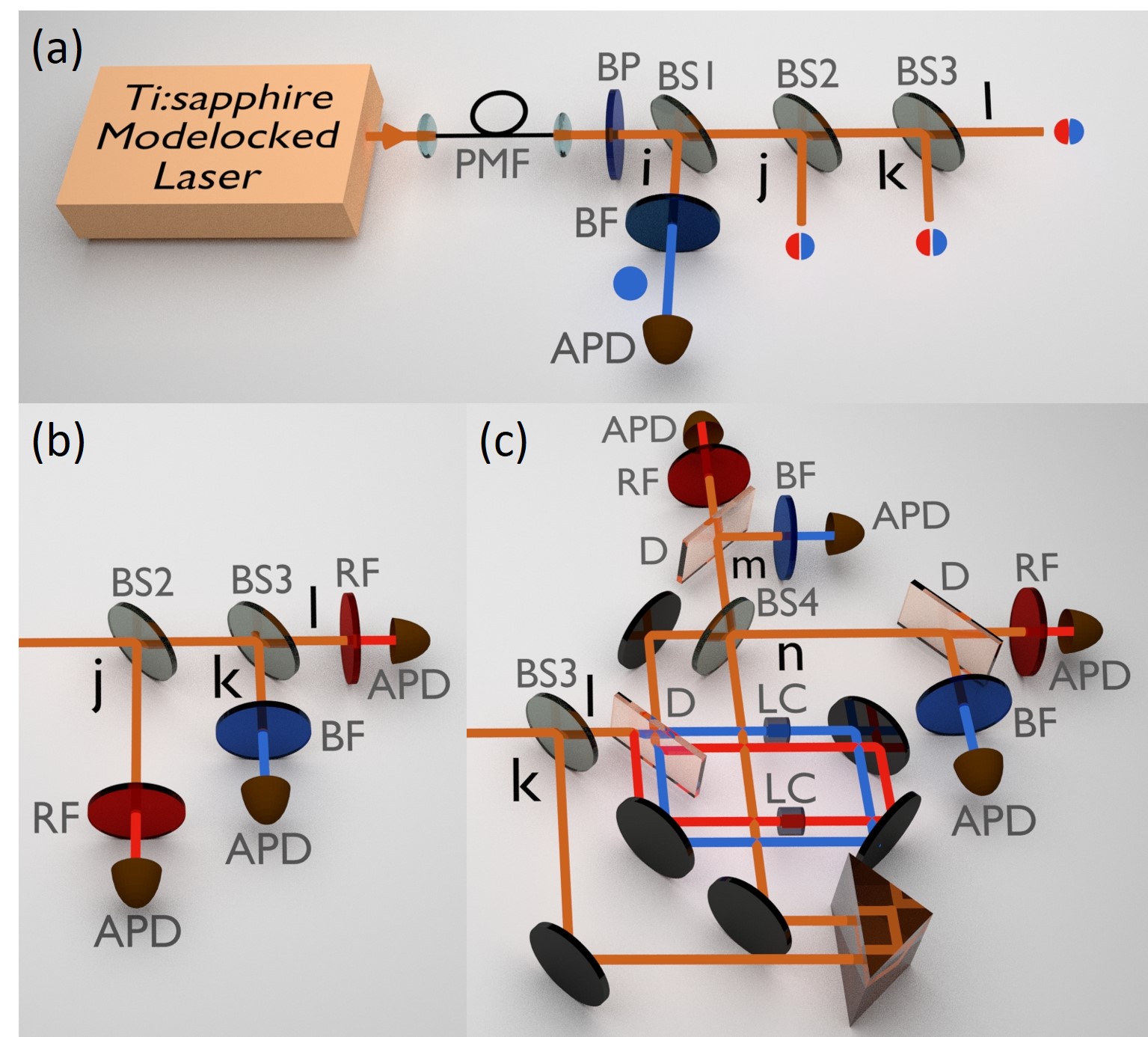}
\caption{(a) Schematic of the experimental setup for generation of the discrete-energy-entangled W state. (b) The setup used to measure coincidences for all possible energy combinations for the three channels $j$, $k$ and $l$, with the filters for measuring the $\ket{Rj,Bk,Rl}$ term shown as an example. The filters are chosen to pass either blue or red photons at each channel for the different projection measurements. (c) Nested interferometers used to measure the reduced density matrix. PMF, polarization-maintaining fiber; BP, bandpass filter; BS1-4, beamsplitter; LC, liquid crystal; D, dichroic mirror; RF, red filter; BF, blue filter; APD, avalanche photodiode.
\label{figure1}}
\end{figure}

The  quantum state of photons generated in optical fibers through SFWM can be written as \cite{Menotti:2016}:

\begin{equation}
\begin{aligned}
\ket{\Psi}=&\left(1-\mathcal{O}(|\beta|^2)\right)\ket{\mathrm{vac}}+ \beta a^{\dagger}_Ba^{\dagger}_R\ket{\mathrm{vac}}\\&+\frac{\beta^2}{2}(a^{\dagger}_B)^2(a^{\dagger}_R)^2\ket{\mathrm{vac}}+\cdots,
\label{eq:psi}
\end{aligned}
\end{equation}
where $|\beta|^2$ is the probability of generating a photon pair, and $a^{\dagger}_B$ and $a^{\dagger}_R$ are the creation operators for signal (blue-detuned from the pump, referred to as blue or $B$) and idler (red or $R$) photons.  We have truncated the expansion at second order, which corresponds to the creation of two
photon pairs. While multi-pair generation is undesirable for photon-pair sources, it can be used to produce multipartite entanglement; compared to using two independent photon-pair sources, using one optical fiber to generate multipartite entanglement is much easier to implement. while still having comparable generation efficiencies. After propagating through a series of beamsplitters, shown in Fig.~\ref{figure1}(a), the state \eqref{eq:psi} becomes \cite{Menotti:2016}:
\begin{equation}\label{state4}
\ket{\Psi}=\alpha\ket{\Psi_0}+\gamma\frac{1}{\sqrt{2}}\left[\ket{B}_i\otimes \ket{W_1}+\ket{R}_i\otimes \ket{W_2}\right],
\end{equation}
where $\ket{\Psi_0}$ encompasses all the terms not leading to a single photon per output channel, $|\alpha|^2 + |\gamma|^2 = 1$,
\begin{eqnarray}\label{eq:W1}
\ket{W_1}=&\frac{1}{\sqrt{3}}(\ket{RRB}_{jkl}+\ket{RBR}_{jkl}+\ket{BRR}_{jkl}),\\
\ket{W_2}=&\frac{1}{\sqrt{3}}(\ket{BBR}_{jkl}+\ket{BRB}_{jkl}+\ket{RBB}_{jkl}),
\label{eq:W2}
\end{eqnarray}
and $i$, $j$, $k$, $l$ indicate channels. The second term of Eq. \eqref{state4} can be taken as the relevant quantum state, assuming that the probability of generating more than two pairs at a time is negligible, and that events associated with single pairs, as well as those without a single photon per channel, can be eliminated through the post-selection of four-fold coincidences. Depending on the detected energy of the photon exiting the first channel $i$, the photons exiting the other three channels $j$, $k$, and $l$ are in one of the W states, $\ket{W_1}$ or $\ket{W_2}$. Note that in general among the three terms  in \eqref{eq:W1} or \eqref{eq:W2}  there will be relative phases associated with the different distances of the source from the three detectors (See Supplemental Material). To retain the coherence of the superposition two conditions must be met: (i) a coherence condition, which requires a typical path length difference $\Delta L$ smaller than the coherence length of each photon, i.e. $\Delta L\ll 2\pi/\Delta \kappa$, where $\Delta \kappa$ is the fluctuation of the wave vector difference determined by the bandwidths of the blue and red photons; (ii) a stability condition, which requires $\delta (\Delta L)\ll 2\pi/{\Delta k_0}$, where $\delta (\Delta L)$ is the fluctuation of the path length difference, and $\Delta k_0 \equiv k_B - k_R$.

A schematic of our experimental setup is shown in Fig.~\ref{figure1}(a). A $1\,\text{m}$ polarization-maintaining optical fiber (PMF, PM780-HP) is pumped by a train of $\sim\!\!100\,\text{fs}$ pulses with center wavelength $810\,\text{nm}$ and repetition rate $80\,\text{MHz}$ from a modelocked Ti:sapphire laser. Pairs of sideband photons are created through SFWM: signal photons at $694\,\text{nm}$ and idler photons at $975\,\text{nm}$. The reflection/transmission ratios of the beamsplitters BS1-3 are chosen to be 25/75, 33/67, and 50/50, respectively, resulting in each photon having a probability of $\frac{1}{4}$ to arrive at each channel, maximizing the generation efficiency of the W state.  Avalanche photodiodes (APDs, SPCM-AQ4C) are placed in each channel to detect the photons. A bandpass filter that only transmits blue is placed in front of detector $i$, such that a four-fold coincidence event corresponds to the post-selection of the state $\ket{W_1}$ for the remaining three photons.
The estimated photon-pair generation efficiency is about 0.02 per pulse for the minimal power we use in the experiment, which gives relatively low higher-order photon-pair generation but enough for detecting two-pair generation. The coupling efficiency and detector efficiency are $80\%$ ($75\%$) and $55\%$ ($11\%$) for signal (idler) photons. Note that the four-fold coincidence rates are already shown in Fig.~2(b).

\begin{figure}[h]

\includegraphics[width=8cm]{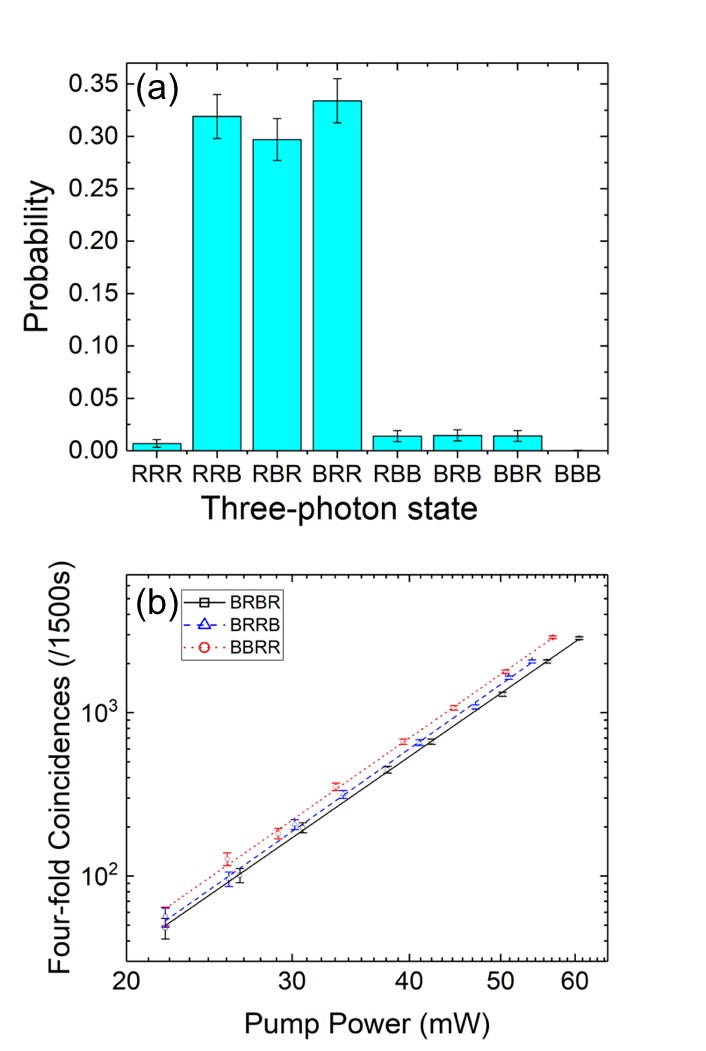}
\caption{(a) Measured probabilities of all possible energy combinations. $\ket{RRB}_{jkl}$, $\ket{RBR}_{jkl}$ and $\ket{BRR}_{jkl}$ are dominant components with about $\frac{1}{3}$ probability, which is as expected for the $\ket{W_1}$ state. (b) Logarithmic plot of four-fold coincidence measurements as a function of pump power. The black, blue and red lines are linear least-squares fits for states $\ket{RBR}_{jkl}$, $\ket{RRB}_{jkl}$ and $\ket{BRR}_{jkl}$, with slopes of  $4.02\,\pm\,0.04$, \,$4.06\pm0.08$ and $4.00\pm0.03$, respectively.}
\label{figure2}

\end{figure}

With moderate pump power ($\sim20\,\text{mW}$), we first measure four-fold coincidences  involving a blue photon in channel $i$  and all possible combinations of energies for the other three channels, namely $\ket{RRR}_{jkl}$, $\ket{RRB}_{jkl}$, $\ket{RBR}_{jkl}$, $\ket{BRR}_{jkl}$, $\ket{RBB}_{jkl}$, $\ket{BRB}_{jkl}$, $\ket{BBR}_{jkl}$, and $\ket{BBB}_{jkl}$. Figure ~\ref{figure1}(b) shows the setup corresponding to the combination $\ket{RBR}_{jkl}$, with bandpass filters used to determine if blue or red photons are detected in each channel. The extracted probabilities are plotted in Fig.~\ref{figure2}(a), in which the three terms $\ket{RRB}_{jkl}$, $\ket{RBR}_{jkl}$, and $\ket{BRR}_{jkl}$, associated with the state $\ket{W_1}$, are dominant with approximately $33\%$ probability each. The other terms are small and come from higher-order pair generation and Raman noise; the lack of counts for the $\ket{BBB}_{jkl}$ term indicates there are no contributions from dark counts. Figure~\ref{figure2}(b) shows the four-fold coincidence counts for the three dominant terms as a function of the pump power. Least-squares fits show that, as expected, the four-fold coincidences scale with the fourth power of the pump power, which confirm these terms arise from two-pair generation. The imbalance between these three terms arises from the imperfection of the beamsplitters.

These measurements are not sufficient to completely characterize the post-selected three photon state. Indeed, the appropriate incoherent mixture of $\ket{RBR}_{jkl}$, $\ket{RRB}_{jkl}$ and $\ket{BRR}_{jkl}$ would lead to the same results as the state $\ket{W_1}$ of Eq.~\ref{eq:W1}.\\ 

To reconstruct the density matrix of the generated state we make use of a strategy based on the so-called "reduced density matrix approach". It  was recently shown that some classes of multipartite entangled states, including three-photon W states, can be determined by constructing a set of reduced density matrices \cite{Xin2017}. This approach was tested experimentally via nuclear magnetic resonance in a molecular sample, where the high fidelities between the results of full quantum state tomography and those deduced from
two-particle reduced density matrices were found to be robust against experimental noise \cite{Xin2017}. Here we use this method to characterize our photonic state. In our case one needs to reconstruct the three reduced density matrices associated with the detection of a red photon in one of the channels $j$, $k$, or $l$ following the detection of a blue photon in channel $i$.

\begin{figure}[h]

\includegraphics[width=8cm]{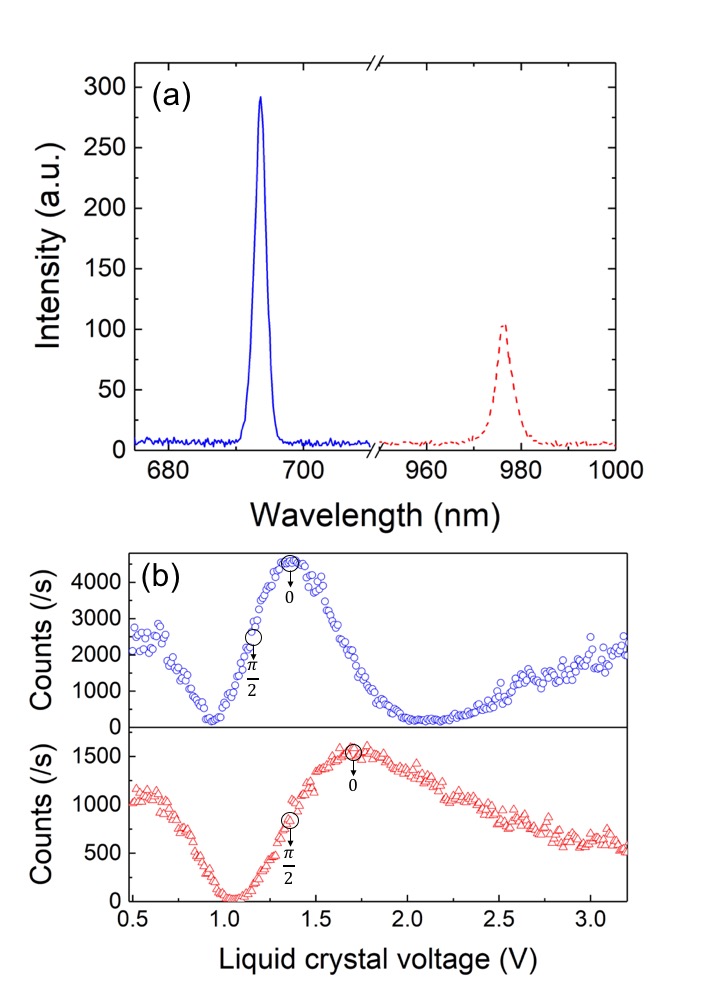}
\caption{(a) Single photon spectrum of blue (blue solid line) and red (red dashed line) photons. (b) Single photon interference counts of blue (blue circles) and red (red triangles)  photons  after the interferometers as a function of voltage across the liquid crystals. The black circles with arrows indicate the voltages across the liquid crystals that are chosen for phases $\phi_B$ and $\phi_R$.}
\label{figure3}

\end{figure}

For example, in the case of a detection of a red photon in channel $j$, the generic  density operator $\rho^{\mathrm{I}}$ describing the photons exiting from channels $k$ and $l$ for experimental runs involving two photons in those channels (see Fig.~\ref{figure1}(c) and the Supplemental Material) is
\begin{equation}
\rho^\mathrm{I}=\sum_{a,b=1}^{10}\ket{a}\bra{a}\rho^{\mathrm{I}}\ket{b}\bra{b}\equiv \sum_{a,b=1}^{10}\ket{a}\rho^{\mathrm{I}}_{ab}\bra{b},\label{state}
\end{equation}
where $\ket{1}=\ket{RR,0}_{kl}$, $\ket{2}=\ket{R,R}_{kl}$, $\ket{3}=\ket{0,RR}_{kl}$, $\ket{4}=\ket{RB,0}_{kl}$, $\ket{5}=\ket{R,B}_{kl}$, $\ket{6}=\ket{B,R}_{kl}$, $\ket{7}=\ket{0,RB}_{kl}$, $\ket{8}=\ket{BB,0}_{kl}$, $\ket{9}=\ket{B,B}_{kl}$, and $\ket{10}=\ket{0,BB}_{kl}$. To determine the matrix elements $\rho^{\mathrm{I}}_{ab}$, a traditional quantum state tomography approach would involve projections on linear combinations of the states of different energies \cite{PhysRevLett.103.253601}, which for the large energy differences of our source would require frequency conversion \cite{ Kues2017, Imany:18}. 
However, in our case we can take advantage of prior information on our source: 1) photons are emitted in pairs by non-degenerate SFWM, and 2) the probability for higher-order pair generation is small. Relying on energy conservation, we thus set to zero all the elements corresponding to photon pairs in which both photons have the same energy (i.e. $\ket{1},\ket{2},\ket{3},\ket{8},\ket{9},\ket{10}$). In our case, this procedure leads to a systematic error of about 3\% given the coincidence-to-accidentals ratio of 30 (see Fig.~\ref{figure2}(a)). The resulting matrix corresponds to a particular class of states that belong to a much smaller Hilbert space and can be characterized through a sequence of linear measurements (see Supplemental Material). We also notice that, since we demonstrate the generation of an energy-entangled $W_1$ state in post-selection on four-fold coincidences, to reconstruct the state of photons exiting different channels we only need to determine the elements $\rho^{\mathrm{I}}_{55}$, $\rho^{\mathrm{I}}_{56}$, and $\rho^{\mathrm{I}}_{66}$.

To determine these elements without frequency conversion, however, we need to go beyond Local Operations and Classical Communications (LOCC) measurements that would be involved in application protocols of W states with parties at separated sites $j$, $k$, $l$. We build two nested interferometers, as shown in Fig.~\ref{figure1}(c). From channel $l$, blue and red photons travel through an offset Sagnac interferometer closed by a dichroic mirror (D). The phases  $\phi_B$ and $\phi_R$ acquired along the blue and red paths, respectively, are controlled independently using two liquid crystal (LC) plates.  The output from the Sagnac interferometer is combined with that from channel $k$ by a 50/50 beamsplitter outputting into two new channels labelled $m$ and $n$.  Dichroic mirrors are placed in channels $m$ and $n$ to separate blue and red photons and facilitate their spectral analysis using four detectors. Finally, bandpass filters are used to suppress pump and Raman noise before detection, which is performed using avalanche photodiodes.

As mentioned before, observing this discrete-energy entangled W state requires that coherence and stability conditions are met. The spectra of the blue and red photons, shown in Fig.~\ref{figure3}(a), are measured using a single-photon-level spectrometer (Andor Shamrock 500i/iDus 420). The measured bandwidths of $2.1\,\text{nm}$ (blue) and $4.3\,\text{nm}$ (red) allow a value of up to approximately $112\,\mu\text{m}$ for the path length difference to satisfy the coherence condition. The effective path length difference can be identified by checking the single-photon interference of blue and red photons after the interferometers in Fig.~\ref{figure1}(c). Due to differing thicknesses of glass in the paths of the blue and red photons, there is an additional delay for the red photons. We insert a pair of wedges in the path of the blue photons within the Sagnac interferometer to compensate for this delay. The stability condition requires the nested interferometers to be phase-stable with fluctuations in the path length $\delta(\Delta L)\ll 1.7 \mu\,\text{m}$. The use of an offset Sagnac geometry provides stability for the inner interferometer. A reference diode laser is sent through the nested interferometers for active stabilization of the outer interferometer. The phases $\phi_B$ and $\phi_R$ are determined through single photon interference of blue and red photons as the voltage across the liquid crystals is tuned.

By means of coincidence measurements on photons exiting channels $m$ and $n$ with phases $\phi_B$ and $\phi_R$ set to $0$ or $\pi/2$ we find \cite{sm} that
\begin{widetext}
\begin{equation}
\begin{aligned}
\text{Re}(\rho^{\mathrm{I}}_{56})=&\frac{1}{2}\left[1-P_{BR}^{(m,n)}(0,0)-P_{RB}^{(m,n)}(0,0)-P_{BR}^{(m,n)}\left(\frac{\pi}{2},\frac{\pi}{2}\right)-P_{RB}^{(m,n)}\left(\frac{\pi}{2},\frac{\pi}{2}\right)\right],\\
\text{Im}(\rho^{\mathrm{I}}_{56})=&\frac{1}{2}\left[P_{BR}^{(m,n)}\left(0,\frac{\pi}{2}\right)+P_{RB}^{(m,n)}\left(0,\frac{\pi}{2}\right)-P_{BR}^{(m,n)}\left(\frac{\pi}{2},0\right)-P_{RB}^{(m,n)}\left(\frac{\pi}{2},0\right)\right],\label{density}
\end{aligned}
\end{equation}
\end{widetext}
where $P_{BR}^{(m,n)}(\phi_B,\phi_R)$ is the probability of a coincidence detection of a blue and a red photon exiting channels $m$ and $n$ when phases $\phi_{B}$ and $\phi_{R}$ are applied, respectively. The circled voltages in Fig.~\ref{figure3}(b) are the experimental set points for $0$ or $\pi/2$ phase as needed in \eqref{density}, which is derived assuming zero phase difference between the two arms of the outer interferometer. Finally, $\rho^{\mathrm{I}}_{55}$ and $\rho^{\mathrm{I}}_{66}$  are measured by simply excluding the last 50/50 beamsplitter (BS4 in Fig.~\ref{figure1}(c)). A full characterization of the tripartite W state is then obtained by iterating the entire procedure on photons exiting from channels $j$\&$l$ and channels $k$\&$j$. In all cases, we post-select upon the detection of a blue photon in $i$ and a red photon in the remaining channel.

\begin{figure}[!h]
\includegraphics[width=8cm]{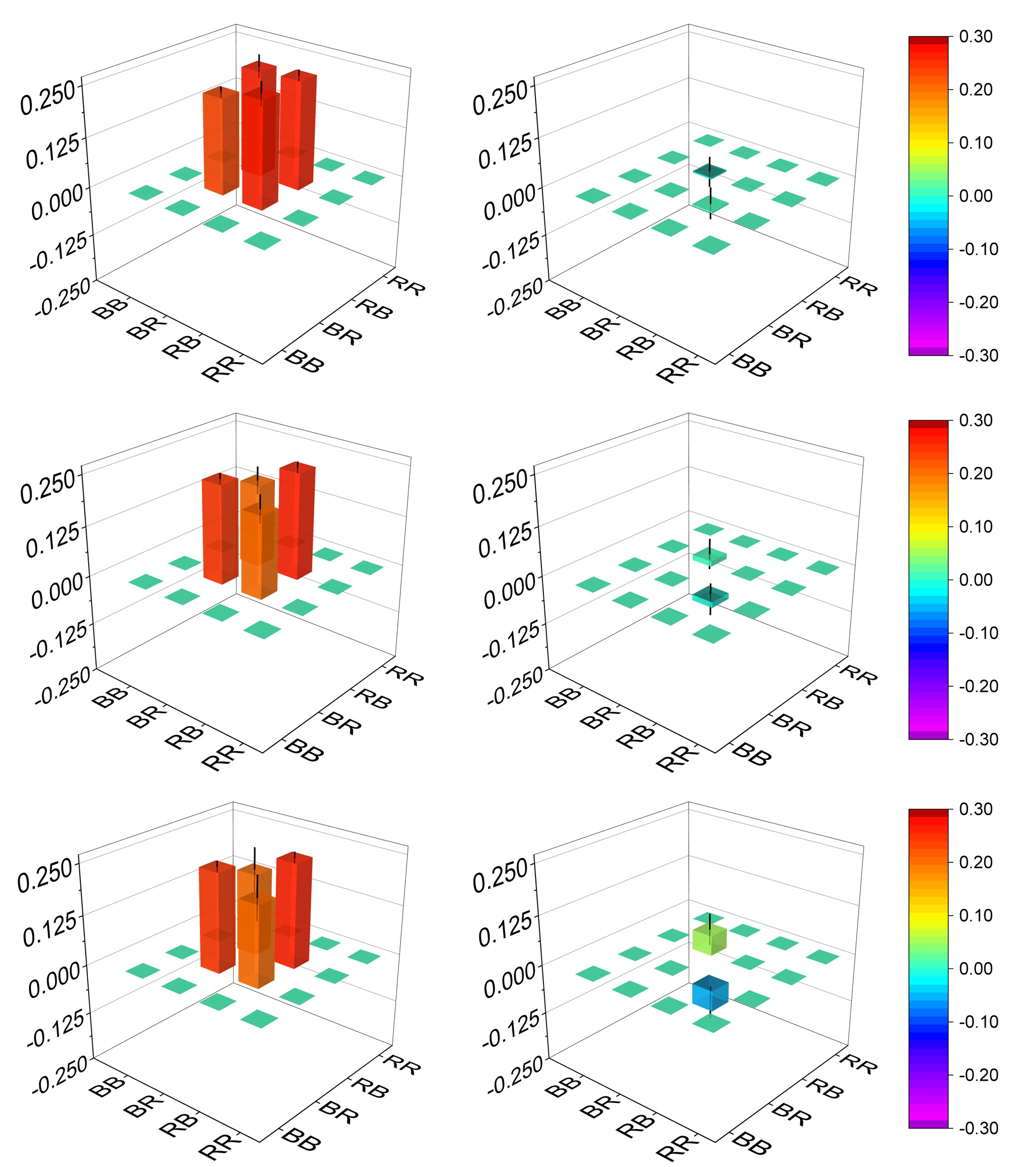}
\caption{The real (left) and imaginary (right) parts of the reconstructed reduced density matrices for $k$\&$l$ (top), $j$\&$l$ (middle) and $k$\&$j$ (bottom) channels. The error bars represent the uncertainty due to phase instability and statistical noise.}
\label{figure4}
\end{figure}

The real and imaginary parts of the reconstructed reduced density matrices describing the photon pairs exiting from different channels upon the detection of a red photon in channel $j$, $k$, or $l$ are shown in Fig.~\ref{figure4}. Please note that we did not subtract accidentals in any of the measured terms. The uncertainties characterizing both the phase instability and the Poissonian statistics of the counts are indicated by the error bars.  From these results we can reconstruct the density matrix of the generated state, and we confirm that this is the discrete-energy-entangled W state $\ket{W_1}$ in \eqref{eq:W1} with an estimated fidelity of $92\%\pm6\%$ and an estimated purity of $90\%\pm10\%$ (see Supplemental Material for the W state density matrix). The fidelity and purity are overestimated due to small but finite counts in some of the zeroed elements; based on Fig.~2(a), the contribution of this systematic error is less than the statistical error of the measured values.


In summary, we have successfully observed the generation of a discrete-energy-entangled W state produced in polarization-maintaining fiber via SFWM. We characterized the state without implementing any frequency conversion, taking advantage of previous information on the source,  reduced density matrix strategy, and experiments which mix the output of different channels. The reconstructed reduced density matrix shows good fidelity to the expected state. With its robustness against noise and loss, we anticipate the discrete-energy-entangled W state and this characterization method to find useful applications in future quantum information protocols. BF and VOL thank D.~G.~England and J.~P.~W.~Maclean for helpful discussion and suggesting the offset Sagnac geometry. VOL would like to thank Offir Cohen for helpful discussions. This work is supported in part by NSF Grant Nos. 1521110, 1640968, and 1806572.


\begin{thebibliography}{36}%
\makeatletter
\providecommand \@ifxundefined [1]{%
 \@ifx{#1\undefined}
}%
\providecommand \@ifnum [1]{%
 \ifnum #1\expandafter \@firstoftwo
 \else \expandafter \@secondoftwo
 \fi
}%
\providecommand \@ifx [1]{%
 \ifx #1\expandafter \@firstoftwo
 \else \expandafter \@secondoftwo
 \fi
}%
\providecommand \natexlab [1]{#1}%
\providecommand \enquote  [1]{``#1''}%
\providecommand \bibnamefont  [1]{#1}%
\providecommand \bibfnamefont [1]{#1}%
\providecommand \citenamefont [1]{#1}%
\providecommand \href@noop [0]{\@secondoftwo}%
\providecommand \href [0]{\begingroup \@sanitize@url \@href}%
\providecommand \@href[1]{\@@startlink{#1}\@@href}%
\providecommand \@@href[1]{\endgroup#1\@@endlink}%
\providecommand \@sanitize@url [0]{\catcode `\\12\catcode `\$12\catcode
  `\&12\catcode `\#12\catcode `\^12\catcode `\_12\catcode `\%12\relax}%
\providecommand \@@startlink[1]{}%
\providecommand \@@endlink[0]{}%
\providecommand \url  [0]{\begingroup\@sanitize@url \@url }%
\providecommand \@url [1]{\endgroup\@href {#1}{\urlprefix }}%
\providecommand \urlprefix  [0]{URL }%
\providecommand \Eprint [0]{\href }%
\providecommand \doibase [0]{http://dx.doi.org/}%
\providecommand \selectlanguage [0]{\@gobble}%
\providecommand \bibinfo  [0]{\@secondoftwo}%
\providecommand \bibfield  [0]{\@secondoftwo}%
\providecommand \translation [1]{[#1]}%
\providecommand \BibitemOpen [0]{}%
\providecommand \bibitemStop [0]{}%
\providecommand \bibitemNoStop [0]{.\EOS\space}%
\providecommand \EOS [0]{\spacefactor3000\relax}%
\providecommand \BibitemShut  [1]{\csname bibitem#1\endcsname}%
\let\auto@bib@innerbib\@empty
\bibitem [{\citenamefont {Mermin}(1990)}]{Mermin1990}%
  \BibitemOpen
  \bibfield  {author} {\bibinfo {author} {\bibfnamefont {N.~D.}\ \bibnamefont
  {Mermin}},\ }\href@noop {} {\bibfield  {journal} {\bibinfo  {journal} {Phys.
  Rev. Lett.}\ }\textbf {\bibinfo {volume} {65}},\ \bibinfo {pages} {1838}
  (\bibinfo {year} {1990})}\BibitemShut {NoStop}%
\bibitem [{\citenamefont {Pan}\ \emph {et~al.}(2000)\citenamefont {Pan},
  \citenamefont {Bouwmeester}, \citenamefont {Daniell}, \citenamefont
  {Weinfurter},\ and\ \citenamefont {Zeilinger}}]{Pan2000}%
  \BibitemOpen
  \bibfield  {author} {\bibinfo {author} {\bibfnamefont {J.-W.}\ \bibnamefont
  {Pan}}, \bibinfo {author} {\bibfnamefont {D.}~\bibnamefont {Bouwmeester}},
  \bibinfo {author} {\bibfnamefont {M.}~\bibnamefont {Daniell}}, \bibinfo
  {author} {\bibfnamefont {H.}~\bibnamefont {Weinfurter}}, \ and\ \bibinfo
  {author} {\bibfnamefont {A.}~\bibnamefont {Zeilinger}},\ }\href@noop {}
  {\bibfield  {journal} {\bibinfo  {journal} {Nature}\ }\textbf {\bibinfo
  {volume} {403}},\ \bibinfo {pages} {515} (\bibinfo {year}
  {2000})}\BibitemShut {NoStop}%
\bibitem [{\citenamefont {Collins}\ \emph {et~al.}(2002)\citenamefont
  {Collins}, \citenamefont {Gisin}, \citenamefont {Popescu}, \citenamefont
  {Roberts},\ and\ \citenamefont {Scarani}}]{Collins2002}%
  \BibitemOpen
  \bibfield  {author} {\bibinfo {author} {\bibfnamefont {D.}~\bibnamefont
  {Collins}}, \bibinfo {author} {\bibfnamefont {N.}~\bibnamefont {Gisin}},
  \bibinfo {author} {\bibfnamefont {S.}~\bibnamefont {Popescu}}, \bibinfo
  {author} {\bibfnamefont {D.}~\bibnamefont {Roberts}}, \ and\ \bibinfo
  {author} {\bibfnamefont {V.}~\bibnamefont {Scarani}},\ }\href@noop {}
  {\bibfield  {journal} {\bibinfo  {journal} {Phys. Rev. Lett.}\ }\textbf
  {\bibinfo {volume} {88}},\ \bibinfo {pages} {170405} (\bibinfo {year}
  {2002})}\BibitemShut {NoStop}%
\bibitem [{\citenamefont {Cabello}(2002)}]{Cabello2002}%
  \BibitemOpen
  \bibfield  {author} {\bibinfo {author} {\bibfnamefont {A.}~\bibnamefont
  {Cabello}},\ }\href {\doibase 10.1103/PhysRevA.65.032108} {\bibfield
  {journal} {\bibinfo  {journal} {Phys. Rev. A}\ }\textbf {\bibinfo {volume}
  {65}},\ \bibinfo {pages} {032108} (\bibinfo {year} {2002})}\BibitemShut
  {NoStop}%
\bibitem [{\citenamefont {Pan}\ \emph {et~al.}(2012)\citenamefont {Pan},
  \citenamefont {Chen}, \citenamefont {Lu}, \citenamefont {Weinfurter},
  \citenamefont {Zeilinger},\ and\ \citenamefont
  {{\stackon[1pt]{Z}{.}}ukowski}}]{Pan2012}%
  \BibitemOpen
  \bibfield  {author} {\bibinfo {author} {\bibfnamefont {J.-W.}\ \bibnamefont
  {Pan}}, \bibinfo {author} {\bibfnamefont {Z.-B.}\ \bibnamefont {Chen}},
  \bibinfo {author} {\bibfnamefont {C.-Y.}\ \bibnamefont {Lu}}, \bibinfo
  {author} {\bibfnamefont {H.}~\bibnamefont {Weinfurter}}, \bibinfo {author}
  {\bibfnamefont {A.}~\bibnamefont {Zeilinger}}, \ and\ \bibinfo {author}
  {\bibfnamefont {M.}~\bibnamefont {{\stackon[1pt]{Z}{.}}ukowski}},\
  }\href@noop {} {\bibfield  {journal} {\bibinfo  {journal} {Reviews of Modern
  Physics}\ }\textbf {\bibinfo {volume} {84}},\ \bibinfo {pages} {777}
  (\bibinfo {year} {2012})}\BibitemShut {NoStop}%
\bibitem [{\citenamefont {Schwaiger}\ \emph {et~al.}(2015)\citenamefont
  {Schwaiger}, \citenamefont {Sauerwein}, \citenamefont {Cuquet}, \citenamefont
  {de~Vicente},\ and\ \citenamefont {Kraus}}]{Schwaiger2015}%
  \BibitemOpen
  \bibfield  {author} {\bibinfo {author} {\bibfnamefont {K.}~\bibnamefont
  {Schwaiger}}, \bibinfo {author} {\bibfnamefont {D.}~\bibnamefont
  {Sauerwein}}, \bibinfo {author} {\bibfnamefont {M.}~\bibnamefont {Cuquet}},
  \bibinfo {author} {\bibfnamefont {J.}~\bibnamefont {de~Vicente}}, \ and\
  \bibinfo {author} {\bibfnamefont {B.}~\bibnamefont {Kraus}},\ }\href@noop {}
  {\bibfield  {journal} {\bibinfo  {journal} {Phys. Rev. Lett.}\ }\textbf
  {\bibinfo {volume} {115}},\ \bibinfo {pages} {150502} (\bibinfo {year}
  {2015})}\BibitemShut {NoStop}%
\bibitem [{\citenamefont {Malik}\ \emph {et~al.}(2016)\citenamefont {Malik},
  \citenamefont {Erhard}, \citenamefont {Huber}, \citenamefont {Krenn},
  \citenamefont {Fickler},\ and\ \citenamefont {Zeilinger}}]{Malik2016}%
  \BibitemOpen
  \bibfield  {author} {\bibinfo {author} {\bibfnamefont {M.}~\bibnamefont
  {Malik}}, \bibinfo {author} {\bibfnamefont {M.}~\bibnamefont {Erhard}},
  \bibinfo {author} {\bibfnamefont {M.}~\bibnamefont {Huber}}, \bibinfo
  {author} {\bibfnamefont {M.}~\bibnamefont {Krenn}}, \bibinfo {author}
  {\bibfnamefont {R.}~\bibnamefont {Fickler}}, \ and\ \bibinfo {author}
  {\bibfnamefont {A.}~\bibnamefont {Zeilinger}},\ }\href@noop {} {\bibfield
  {journal} {\bibinfo  {journal} {Nat. Photonics}\ }\textbf {\bibinfo {volume}
  {10}},\ \bibinfo {pages} {248} (\bibinfo {year} {2016})}\BibitemShut
  {NoStop}%
\bibitem [{\citenamefont {Horodecki}\ \emph {et~al.}(2009)\citenamefont
  {Horodecki}, \citenamefont {Horodecki}, \citenamefont {Horodecki},\ and\
  \citenamefont {Horodecki}}]{Horodecki2009}%
  \BibitemOpen
  \bibfield  {author} {\bibinfo {author} {\bibfnamefont {R.}~\bibnamefont
  {Horodecki}}, \bibinfo {author} {\bibfnamefont {P.}~\bibnamefont
  {Horodecki}}, \bibinfo {author} {\bibfnamefont {M.}~\bibnamefont
  {Horodecki}}, \ and\ \bibinfo {author} {\bibfnamefont {K.}~\bibnamefont
  {Horodecki}},\ }\href@noop {} {\bibfield  {journal} {\bibinfo  {journal}
  {Reviews of Modern Physics}\ }\textbf {\bibinfo {volume} {81}},\ \bibinfo
  {pages} {865} (\bibinfo {year} {2009})}\BibitemShut {NoStop}%
\bibitem [{\citenamefont {Greenberger}\ \emph {et~al.}(1990)\citenamefont
  {Greenberger}, \citenamefont {Horne}, \citenamefont {Shimony},\ and\
  \citenamefont {Zeilinger}}]{Greenberger1990}%
  \BibitemOpen
  \bibfield  {author} {\bibinfo {author} {\bibfnamefont {D.~M.}\ \bibnamefont
  {Greenberger}}, \bibinfo {author} {\bibfnamefont {M.~A.}\ \bibnamefont
  {Horne}}, \bibinfo {author} {\bibfnamefont {A.}~\bibnamefont {Shimony}}, \
  and\ \bibinfo {author} {\bibfnamefont {A.}~\bibnamefont {Zeilinger}},\
  }\href@noop {} {\bibfield  {journal} {\bibinfo  {journal} {Am. J. Phys.}\
  }\textbf {\bibinfo {volume} {58}},\ \bibinfo {pages} {12} (\bibinfo {year}
  {1990})}\BibitemShut {NoStop}%
\bibitem [{\citenamefont {Bouwmeester}\ \emph {et~al.}(1999)\citenamefont
  {Bouwmeester}, \citenamefont {Pan}, \citenamefont {Daniell}, \citenamefont
  {Weinfurter},\ and\ \citenamefont {Zeilinger}}]{PhysRevLett.82.1345}%
  \BibitemOpen
  \bibfield  {author} {\bibinfo {author} {\bibfnamefont {D.}~\bibnamefont
  {Bouwmeester}}, \bibinfo {author} {\bibfnamefont {J.-W.}\ \bibnamefont
  {Pan}}, \bibinfo {author} {\bibfnamefont {M.}~\bibnamefont {Daniell}},
  \bibinfo {author} {\bibfnamefont {H.}~\bibnamefont {Weinfurter}}, \ and\
  \bibinfo {author} {\bibfnamefont {A.}~\bibnamefont {Zeilinger}},\ }\href@noop
  {} {\bibfield  {journal} {\bibinfo  {journal} {Phys. Rev. Lett.}\ }\textbf
  {\bibinfo {volume} {82}},\ \bibinfo {pages} {1345} (\bibinfo {year}
  {1999})}\BibitemShut {NoStop}%
\bibitem [{\citenamefont {D{\"u}r}\ \emph {et~al.}(2000)\citenamefont
  {D{\"u}r}, \citenamefont {Vidal},\ and\ \citenamefont
  {Cirac}}]{PhysRevA.62.062314}%
  \BibitemOpen
  \bibfield  {author} {\bibinfo {author} {\bibfnamefont {W.}~\bibnamefont
  {D{\"u}r}}, \bibinfo {author} {\bibfnamefont {G.}~\bibnamefont {Vidal}}, \
  and\ \bibinfo {author} {\bibfnamefont {J.~I.}\ \bibnamefont {Cirac}},\
  }\href@noop {} {\bibfield  {journal} {\bibinfo  {journal} {Phys. Rev. A}\
  }\textbf {\bibinfo {volume} {62}},\ \bibinfo {pages} {{062314}} (\bibinfo
  {year} {2000})}\BibitemShut {NoStop}%
\bibitem [{\citenamefont {Barnea}\ \emph {et~al.}(2015)\citenamefont {Barnea},
  \citenamefont {P{\"u}tz}, \citenamefont {Brask}, \citenamefont {Brunner},
  \citenamefont {Gisin},\ and\ \citenamefont {Liang}}]{Barnea2015}%
  \BibitemOpen
  \bibfield  {author} {\bibinfo {author} {\bibfnamefont {T.~J.}\ \bibnamefont
  {Barnea}}, \bibinfo {author} {\bibfnamefont {G.}~\bibnamefont {P{\"u}tz}},
  \bibinfo {author} {\bibfnamefont {J.~B.}\ \bibnamefont {Brask}}, \bibinfo
  {author} {\bibfnamefont {N.}~\bibnamefont {Brunner}}, \bibinfo {author}
  {\bibfnamefont {N.}~\bibnamefont {Gisin}}, \ and\ \bibinfo {author}
  {\bibfnamefont {Y.-C.}\ \bibnamefont {Liang}},\ }\href {\doibase
  10.1103/PhysRevA.91.032108} {\bibfield  {journal} {\bibinfo  {journal} {Phys.
  Rev. A}\ }\textbf {\bibinfo {volume} {91}},\ \bibinfo {pages} {032108}
  (\bibinfo {year} {2015})}\BibitemShut {NoStop}%
\bibitem [{\citenamefont {Sohbi}\ \emph {et~al.}(2015)\citenamefont {Sohbi},
  \citenamefont {Zaquine}, \citenamefont {Diamanti},\ and\ \citenamefont
  {Markham}}]{Sohbi2015}%
  \BibitemOpen
  \bibfield  {author} {\bibinfo {author} {\bibfnamefont {A.}~\bibnamefont
  {Sohbi}}, \bibinfo {author} {\bibfnamefont {I.}~\bibnamefont {Zaquine}},
  \bibinfo {author} {\bibfnamefont {E.}~\bibnamefont {Diamanti}}, \ and\
  \bibinfo {author} {\bibfnamefont {D.}~\bibnamefont {Markham}},\ }\href
  {\doibase 10.1103/PhysRevA.91.022101} {\bibfield  {journal} {\bibinfo
  {journal} {Phys. Rev. A}\ }\textbf {\bibinfo {volume} {91}},\ \bibinfo
  {pages} {022101} (\bibinfo {year} {2015})}\BibitemShut {NoStop}%
\bibitem [{\citenamefont {Sen(De)}\ \emph {et~al.}(2003)\citenamefont
  {Sen(De)}, \citenamefont {Sen}, \citenamefont {Wie{\'s}niak}, \citenamefont
  {Kaszlikowski},\ and\ \citenamefont
  {{\stackon[1pt]{Z}{.}}ukowski}}]{SenDe:PhysicalReviewA:2003}%
  \BibitemOpen
  \bibfield  {author} {\bibinfo {author} {\bibfnamefont {A.}~\bibnamefont
  {Sen(De)}}, \bibinfo {author} {\bibfnamefont {U.}~\bibnamefont {Sen}},
  \bibinfo {author} {\bibfnamefont {M.}~\bibnamefont {Wie{\'s}niak}}, \bibinfo
  {author} {\bibfnamefont {D.}~\bibnamefont {Kaszlikowski}}, \ and\ \bibinfo
  {author} {\bibfnamefont {M.}~\bibnamefont {{\stackon[1pt]{Z}{.}}ukowski}},\
  }\href@noop {} {\bibfield  {journal} {\bibinfo  {journal} {Phys. Rev. A}\
  }\textbf {\bibinfo {volume} {68}} (\bibinfo {year} {2003})}\BibitemShut
  {NoStop}%
\bibitem [{\citenamefont {Murao}\ \emph {et~al.}(1999)\citenamefont {Murao},
  \citenamefont {Jonathan}, \citenamefont {Plenio},\ and\ \citenamefont
  {Vedral}}]{Murao1999}%
  \BibitemOpen
  \bibfield  {author} {\bibinfo {author} {\bibfnamefont {M.}~\bibnamefont
  {Murao}}, \bibinfo {author} {\bibfnamefont {D.}~\bibnamefont {Jonathan}},
  \bibinfo {author} {\bibfnamefont {M.~B.}\ \bibnamefont {Plenio}}, \ and\
  \bibinfo {author} {\bibfnamefont {V.}~\bibnamefont {Vedral}},\ }\href@noop {}
  {\bibfield  {journal} {\bibinfo  {journal} {Phys. Rev. A}\ }\textbf {\bibinfo
  {volume} {59}},\ \bibinfo {pages} {156} (\bibinfo {year} {1999})}\BibitemShut
  {NoStop}%
\bibitem [{\citenamefont {Shi}\ and\ \citenamefont {Tomita}(2002)}]{Shi2002}%
  \BibitemOpen
  \bibfield  {author} {\bibinfo {author} {\bibfnamefont {B.-S.}\ \bibnamefont
  {Shi}}\ and\ \bibinfo {author} {\bibfnamefont {A.}~\bibnamefont {Tomita}},\
  }\href@noop {} {\bibfield  {journal} {\bibinfo  {journal} {Physics Letters
  A}\ }\textbf {\bibinfo {volume} {296}},\ \bibinfo {pages} {161} (\bibinfo
  {year} {2002})}\BibitemShut {NoStop}%
\bibitem [{\citenamefont {Joo}\ \emph {et~al.}(2003)\citenamefont {Joo},
  \citenamefont {Park}, \citenamefont {Oh},\ and\ \citenamefont
  {Kim}}]{Joo2003}%
  \BibitemOpen
  \bibfield  {author} {\bibinfo {author} {\bibfnamefont {J.}~\bibnamefont
  {Joo}}, \bibinfo {author} {\bibfnamefont {Y.-J.}\ \bibnamefont {Park}},
  \bibinfo {author} {\bibfnamefont {S.}~\bibnamefont {Oh}}, \ and\ \bibinfo
  {author} {\bibfnamefont {J.}~\bibnamefont {Kim}},\ }\href@noop {} {\bibfield
  {journal} {\bibinfo  {journal} {New Journal of Physics}\ }\textbf {\bibinfo
  {volume} {5}},\ \bibinfo {pages} {136} (\bibinfo {year} {2003})}\BibitemShut
  {NoStop}%
\bibitem [{\citenamefont {Gorbachev}\ \emph {et~al.}(2003)\citenamefont
  {Gorbachev}, \citenamefont {Trubilko}, \citenamefont {Rodichkina},\ and\
  \citenamefont {Zhiliba}}]{Gorbachev2003}%
  \BibitemOpen
  \bibfield  {author} {\bibinfo {author} {\bibfnamefont {V.}~\bibnamefont
  {Gorbachev}}, \bibinfo {author} {\bibfnamefont {A.}~\bibnamefont {Trubilko}},
  \bibinfo {author} {\bibfnamefont {A.}~\bibnamefont {Rodichkina}}, \ and\
  \bibinfo {author} {\bibfnamefont {A.}~\bibnamefont {Zhiliba}},\ }\href@noop
  {} {\bibfield  {journal} {\bibinfo  {journal} {Physics Letters A}\ }\textbf
  {\bibinfo {volume} {314}},\ \bibinfo {pages} {267} (\bibinfo {year}
  {2003})}\BibitemShut {NoStop}%
\bibitem [{\citenamefont {Agrawal}\ and\ \citenamefont
  {Pati}(2006)}]{Agrawal:PhysicalReviewA:2006}%
  \BibitemOpen
  \bibfield  {author} {\bibinfo {author} {\bibfnamefont {P.}~\bibnamefont
  {Agrawal}}\ and\ \bibinfo {author} {\bibfnamefont {A.}~\bibnamefont {Pati}},\
  }\href@noop {} {\bibfield  {journal} {\bibinfo  {journal} {Phys. Rev. A}\
  }\textbf {\bibinfo {volume} {74}} (\bibinfo {year} {2006})}\BibitemShut
  {NoStop}%
\bibitem [{\citenamefont {Yamamoto}\ \emph {et~al.}(2002)\citenamefont
  {Yamamoto}, \citenamefont {Tamaki}, \citenamefont {Koashi},\ and\
  \citenamefont {Imoto}}]{Yamamoto2002}%
  \BibitemOpen
  \bibfield  {author} {\bibinfo {author} {\bibfnamefont {T.}~\bibnamefont
  {Yamamoto}}, \bibinfo {author} {\bibfnamefont {K.}~\bibnamefont {Tamaki}},
  \bibinfo {author} {\bibfnamefont {M.}~\bibnamefont {Koashi}}, \ and\ \bibinfo
  {author} {\bibfnamefont {N.}~\bibnamefont {Imoto}},\ }\href@noop {}
  {\bibfield  {journal} {\bibinfo  {journal} {Phys. Rev. A}\ }\textbf {\bibinfo
  {volume} {66}} (\bibinfo {year} {2002})}\BibitemShut {NoStop}%
\bibitem [{\citenamefont {Kiesel}\ \emph {et~al.}(2003)\citenamefont {Kiesel},
  \citenamefont {Bourennane}, \citenamefont {Kurtsiefer}, \citenamefont
  {Weinfurter}, \citenamefont {Kaszlikowski}, \citenamefont {Laskowski},\ and\
  \citenamefont {Zukowski}}]{Kiesel2003}%
  \BibitemOpen
  \bibfield  {author} {\bibinfo {author} {\bibfnamefont {N.}~\bibnamefont
  {Kiesel}}, \bibinfo {author} {\bibfnamefont {M.}~\bibnamefont {Bourennane}},
  \bibinfo {author} {\bibfnamefont {C.}~\bibnamefont {Kurtsiefer}}, \bibinfo
  {author} {\bibfnamefont {H.}~\bibnamefont {Weinfurter}}, \bibinfo {author}
  {\bibfnamefont {D.}~\bibnamefont {Kaszlikowski}}, \bibinfo {author}
  {\bibfnamefont {W.}~\bibnamefont {Laskowski}}, \ and\ \bibinfo {author}
  {\bibfnamefont {M.}~\bibnamefont {Zukowski}},\ }\href@noop {} {\bibfield
  {journal} {\bibinfo  {journal} {Journal of Modern Optics}\ }\textbf {\bibinfo
  {volume} {50}},\ \bibinfo {pages} {1131} (\bibinfo {year}
  {2003})}\BibitemShut {NoStop}%
\bibitem [{\citenamefont {Eibl}\ \emph {et~al.}(2004)\citenamefont {Eibl},
  \citenamefont {Kiesel}, \citenamefont {Bourennane}, \citenamefont
  {Kurtsiefer},\ and\ \citenamefont {Weinfurter}}]{Eibl2004}%
  \BibitemOpen
  \bibfield  {author} {\bibinfo {author} {\bibfnamefont {M.}~\bibnamefont
  {Eibl}}, \bibinfo {author} {\bibfnamefont {N.}~\bibnamefont {Kiesel}},
  \bibinfo {author} {\bibfnamefont {M.}~\bibnamefont {Bourennane}}, \bibinfo
  {author} {\bibfnamefont {C.}~\bibnamefont {Kurtsiefer}}, \ and\ \bibinfo
  {author} {\bibfnamefont {H.}~\bibnamefont {Weinfurter}},\ }\href@noop {}
  {\bibfield  {journal} {\bibinfo  {journal} {Phys. Rev. Lett.}\ }\textbf
  {\bibinfo {volume} {92}},\ \bibinfo {pages} {077901} (\bibinfo {year}
  {2004})}\BibitemShut {NoStop}%
\bibitem [{\citenamefont {Mikami}\ \emph {et~al.}(2004)\citenamefont {Mikami},
  \citenamefont {Li},\ and\ \citenamefont {Kobayashi}}]{Mikami2004}%
  \BibitemOpen
  \bibfield  {author} {\bibinfo {author} {\bibfnamefont {H.}~\bibnamefont
  {Mikami}}, \bibinfo {author} {\bibfnamefont {Y.}~\bibnamefont {Li}}, \ and\
  \bibinfo {author} {\bibfnamefont {T.}~\bibnamefont {Kobayashi}},\ }\href@noop
  {} {\bibfield  {journal} {\bibinfo  {journal} {Phys. Rev. A}\ }\textbf
  {\bibinfo {volume} {70}} (\bibinfo {year} {2004})}\BibitemShut {NoStop}%
\bibitem [{\citenamefont {Mikami}\ \emph {et~al.}(2005)\citenamefont {Mikami},
  \citenamefont {Li}, \citenamefont {Fukuoka},\ and\ \citenamefont
  {Kobayashi}}]{PhysRevLett.95.150404}%
  \BibitemOpen
  \bibfield  {author} {\bibinfo {author} {\bibfnamefont {H.}~\bibnamefont
  {Mikami}}, \bibinfo {author} {\bibfnamefont {Y.}~\bibnamefont {Li}}, \bibinfo
  {author} {\bibfnamefont {K.}~\bibnamefont {Fukuoka}}, \ and\ \bibinfo
  {author} {\bibfnamefont {T.}~\bibnamefont {Kobayashi}},\ }\href {\doibase
  10.1103/PhysRevLett.95.150404} {\bibfield  {journal} {\bibinfo  {journal}
  {Phys. Rev. Lett.}\ }\textbf {\bibinfo {volume} {95}},\ \bibinfo {pages}
  {150404} (\bibinfo {year} {2005})}\BibitemShut {NoStop}%
\bibitem [{\citenamefont {Antonelli}\ \emph {et~al.}(2011)\citenamefont
  {Antonelli}, \citenamefont {Shtaif},\ and\ \citenamefont
  {Brodsky}}]{PhysRevLett.106.080404}%
  \BibitemOpen
  \bibfield  {author} {\bibinfo {author} {\bibfnamefont {C.}~\bibnamefont
  {Antonelli}}, \bibinfo {author} {\bibfnamefont {M.}~\bibnamefont {Shtaif}}, \
  and\ \bibinfo {author} {\bibfnamefont {M.}~\bibnamefont {Brodsky}},\ }\href
  {\doibase 10.1103/PhysRevLett.106.080404} {\bibfield  {journal} {\bibinfo
  {journal} {Phys. Rev. Lett.}\ }\textbf {\bibinfo {volume} {106}},\ \bibinfo
  {pages} {080404} (\bibinfo {year} {2011})}\BibitemShut {NoStop}%
\bibitem [{\citenamefont {Menotti}\ \emph {et~al.}(2016)\citenamefont
  {Menotti}, \citenamefont {Maccone}, \citenamefont {Sipe},\ and\ \citenamefont
  {Liscidini}}]{Menotti:2016}%
  \BibitemOpen
  \bibfield  {author} {\bibinfo {author} {\bibfnamefont {M.}~\bibnamefont
  {Menotti}}, \bibinfo {author} {\bibfnamefont {L.}~\bibnamefont {Maccone}},
  \bibinfo {author} {\bibfnamefont {J.~E.}\ \bibnamefont {Sipe}}, \ and\
  \bibinfo {author} {\bibfnamefont {M.}~\bibnamefont {Liscidini}},\ }\href@noop
  {} {\bibfield  {journal} {\bibinfo  {journal} {Phys. Rev. A}\ }\textbf
  {\bibinfo {volume} {94}},\ \bibinfo
  {pages} {013845} (\bibinfo {year} {2016})}\BibitemShut {NoStop}%
\bibitem [{\citenamefont {Huang}\ and\ \citenamefont
  {Kumar}(1992)}]{PhysRevLett.68.2153}%
  \BibitemOpen
  \bibfield  {author} {\bibinfo {author} {\bibfnamefont {J.}~\bibnamefont
  {Huang}}\ and\ \bibinfo {author} {\bibfnamefont {P.}~\bibnamefont {Kumar}},\
  }\href {\doibase 10.1103/PhysRevLett.68.2153} {\bibfield  {journal} {\bibinfo
   {journal} {Phys. Rev. Lett.}\ }\textbf {\bibinfo {volume} {68}},\ \bibinfo
  {pages} {2153} (\bibinfo {year} {1992})}\BibitemShut {NoStop}%
\bibitem [{\citenamefont {Zaske}\ \emph {et~al.}(2012)\citenamefont {Zaske},
  \citenamefont {Lenhard}, \citenamefont {Ke{\ss}ler}, \citenamefont {Kettler},
  \citenamefont {Hepp}, \citenamefont {Arend}, \citenamefont {Albrecht},
  \citenamefont {Schulz}, \citenamefont {Jetter}, \citenamefont {Michler},\
  and\ \citenamefont {Becher}}]{PhysRevLett.109.147404}%
  \BibitemOpen
  \bibfield  {author} {\bibinfo {author} {\bibfnamefont {S.}~\bibnamefont
  {Zaske}}, \bibinfo {author} {\bibfnamefont {A.}~\bibnamefont {Lenhard}},
  \bibinfo {author} {\bibfnamefont {C.~A.}\ \bibnamefont {Ke{\ss}ler}},
  \bibinfo {author} {\bibfnamefont {J.}~\bibnamefont {Kettler}}, \bibinfo
  {author} {\bibfnamefont {C.}~\bibnamefont {Hepp}}, \bibinfo {author}
  {\bibfnamefont {C.}~\bibnamefont {Arend}}, \bibinfo {author} {\bibfnamefont
  {R.}~\bibnamefont {Albrecht}}, \bibinfo {author} {\bibfnamefont {W.-M.}\
  \bibnamefont {Schulz}}, \bibinfo {author} {\bibfnamefont {M.}~\bibnamefont
  {Jetter}}, \bibinfo {author} {\bibfnamefont {P.}~\bibnamefont {Michler}}, \
  and\ \bibinfo {author} {\bibfnamefont {C.}~\bibnamefont {Becher}},\ }\href
  {\doibase 10.1103/PhysRevLett.109.147404} {\bibfield  {journal} {\bibinfo
  {journal} {Phys. Rev. Lett.}\ }\textbf {\bibinfo {volume} {109}},\ \bibinfo
  {pages} {147404} (\bibinfo {year} {2012})}\BibitemShut {NoStop}%
\bibitem [{\citenamefont {McGuinness}\ \emph {et~al.}(2010)\citenamefont
  {McGuinness}, \citenamefont {Raymer}, \citenamefont {McKinstrie},\ and\
  \citenamefont {Radic}}]{PhysRevLett.105.093604}%
  \BibitemOpen
  \bibfield  {author} {\bibinfo {author} {\bibfnamefont {H.~J.}\ \bibnamefont
  {McGuinness}}, \bibinfo {author} {\bibfnamefont {M.~G.}\ \bibnamefont
  {Raymer}}, \bibinfo {author} {\bibfnamefont {C.~J.}\ \bibnamefont
  {McKinstrie}}, \ and\ \bibinfo {author} {\bibfnamefont {S.}~\bibnamefont
  {Radic}},\ }\href {\doibase 10.1103/PhysRevLett.105.093604} {\bibfield
  {journal} {\bibinfo  {journal} {Phys. Rev. Lett.}\ }\textbf {\bibinfo
  {volume} {105}},\ \bibinfo {pages} {093604} (\bibinfo {year}
  {2010})}\BibitemShut {NoStop}%
\bibitem [{\citenamefont {Agha}\ \emph {et~al.}(2012)\citenamefont {Agha},
  \citenamefont {Davan{\c{c}}o}, \citenamefont {Thurston},\ and\ \citenamefont
  {Srinivasan}}]{Agha:12}%
  \BibitemOpen
  \bibfield  {author} {\bibinfo {author} {\bibfnamefont {I.}~\bibnamefont
  {Agha}}, \bibinfo {author} {\bibfnamefont {M.}~\bibnamefont {Davan{\c{c}}o}},
  \bibinfo {author} {\bibfnamefont {B.}~\bibnamefont {Thurston}}, \ and\
  \bibinfo {author} {\bibfnamefont {K.}~\bibnamefont {Srinivasan}},\ }\href
  {\doibase 10.1364/OL.37.002997} {\bibfield  {journal} {\bibinfo  {journal}
  {Opt. Lett.}\ }\textbf {\bibinfo {volume} {37}},\ \bibinfo {pages} {2997}
  (\bibinfo {year} {2012})}\BibitemShut {NoStop}%
\bibitem [{\citenamefont {Grassani}\ \emph {et~al.}(2015)\citenamefont
  {Grassani}, \citenamefont {Azzini}, \citenamefont {Liscidini}, \citenamefont
  {Galli}, \citenamefont {Strain}, \citenamefont {Sorel}, \citenamefont
  {Sipe},\ and\ \citenamefont {Bajoni}}]{Grassani:15}%
  \BibitemOpen
  \bibfield  {author} {\bibinfo {author} {\bibfnamefont {D.}~\bibnamefont
  {Grassani}}, \bibinfo {author} {\bibfnamefont {S.}~\bibnamefont {Azzini}},
  \bibinfo {author} {\bibfnamefont {M.}~\bibnamefont {Liscidini}}, \bibinfo
  {author} {\bibfnamefont {M.}~\bibnamefont {Galli}}, \bibinfo {author}
  {\bibfnamefont {M.~J.}\ \bibnamefont {Strain}}, \bibinfo {author}
  {\bibfnamefont {M.}~\bibnamefont {Sorel}}, \bibinfo {author} {\bibfnamefont
  {J.~E.}\ \bibnamefont {Sipe}}, \ and\ \bibinfo {author} {\bibfnamefont
  {D.}~\bibnamefont {Bajoni}},\ }\href {\doibase 10.1364/OPTICA.2.000088}
  {\bibfield  {journal} {\bibinfo  {journal} {Optica}\ }\textbf {\bibinfo
  {volume} {2}},\ \bibinfo {pages} {88} (\bibinfo {year} {2015})}\BibitemShut
  {NoStop}%
\bibitem [{\citenamefont {Xin}\ \emph {et~al.}(2017)\citenamefont {Xin},
  \citenamefont {Lu}, \citenamefont {Klassen}, \citenamefont {Yu},
  \citenamefont {Ji}, \citenamefont {Chen}, \citenamefont {Ma}, \citenamefont
  {Long}, \citenamefont {Zeng},\ and\ \citenamefont {Laflamme}}]{Xin2017}%
  \BibitemOpen
  \bibfield  {author} {\bibinfo {author} {\bibfnamefont {T.}~\bibnamefont
  {Xin}}, \bibinfo {author} {\bibfnamefont {D.}~\bibnamefont {Lu}}, \bibinfo
  {author} {\bibfnamefont {J.}~\bibnamefont {Klassen}}, \bibinfo {author}
  {\bibfnamefont {N.}~\bibnamefont {Yu}}, \bibinfo {author} {\bibfnamefont
  {Z.}~\bibnamefont {Ji}}, \bibinfo {author} {\bibfnamefont {J.}~\bibnamefont
  {Chen}}, \bibinfo {author} {\bibfnamefont {X.}~\bibnamefont {Ma}}, \bibinfo
  {author} {\bibfnamefont {G.}~\bibnamefont {Long}}, \bibinfo {author}
  {\bibfnamefont {B.}~\bibnamefont {Zeng}}, \ and\ \bibinfo {author}
  {\bibfnamefont {R.}~\bibnamefont {Laflamme}},\ }\href@noop {} {\bibfield
  {journal} {\bibinfo  {journal} {Phys. Rev. Lett.}\ }\textbf {\bibinfo
  {volume} {118}},\ \bibinfo {pages} {020401} (\bibinfo {year}
  {2017})}\BibitemShut {NoStop}%
\bibitem [{\citenamefont {Ramelow}\ \emph {et~al.}(2009)\citenamefont
  {Ramelow}, \citenamefont {Ratschbacher}, \citenamefont {Fedrizzi},
  \citenamefont {Langford},\ and\ \citenamefont
  {Zeilinger}}]{PhysRevLett.103.253601}%
  \BibitemOpen
  \bibfield  {author} {\bibinfo {author} {\bibfnamefont {S.}~\bibnamefont
  {Ramelow}}, \bibinfo {author} {\bibfnamefont {L.}~\bibnamefont
  {Ratschbacher}}, \bibinfo {author} {\bibfnamefont {A.}~\bibnamefont
  {Fedrizzi}}, \bibinfo {author} {\bibfnamefont {N.~K.}\ \bibnamefont
  {Langford}}, \ and\ \bibinfo {author} {\bibfnamefont {A.}~\bibnamefont
  {Zeilinger}},\ }\href {\doibase 10.1103/PhysRevLett.103.253601} {\bibfield
  {journal} {\bibinfo  {journal} {Phys. Rev. Lett.}\ }\textbf {\bibinfo
  {volume} {103}},\ \bibinfo {pages} {253601} (\bibinfo {year}
  {2009})}\BibitemShut {NoStop}%
\bibitem [{\citenamefont {Kues}\ \emph {et~al.}(2017)\citenamefont {Kues},
  \citenamefont {Reimer}, \citenamefont {Roztocki}, \citenamefont {Cort{\'e}s},
  \citenamefont {Sciara}, \citenamefont {Wetzel}, \citenamefont {Zhang},
  \citenamefont {Cino}, \citenamefont {Chu}, \citenamefont {Little},
  \citenamefont {Moss}, \citenamefont {Caspani}, \citenamefont {Aza{\~n}a},\
  and\ \citenamefont {Morandotti}}]{Kues2017}%
  \BibitemOpen
  \bibfield  {author} {\bibinfo {author} {\bibfnamefont {M.}~\bibnamefont
  {Kues}}, \bibinfo {author} {\bibfnamefont {C.}~\bibnamefont {Reimer}},
  \bibinfo {author} {\bibfnamefont {P.}~\bibnamefont {Roztocki}}, \bibinfo
  {author} {\bibfnamefont {L.~R.}\ \bibnamefont {Cort{\'e}s}}, \bibinfo
  {author} {\bibfnamefont {S.}~\bibnamefont {Sciara}}, \bibinfo {author}
  {\bibfnamefont {B.}~\bibnamefont {Wetzel}}, \bibinfo {author} {\bibfnamefont
  {Y.}~\bibnamefont {Zhang}}, \bibinfo {author} {\bibfnamefont
  {A.}~\bibnamefont {Cino}}, \bibinfo {author} {\bibfnamefont {S.~T.}\
  \bibnamefont {Chu}}, \bibinfo {author} {\bibfnamefont {B.~E.}\ \bibnamefont
  {Little}}, \bibinfo {author} {\bibfnamefont {D.~J.}\ \bibnamefont {Moss}},
  \bibinfo {author} {\bibfnamefont {L.}~\bibnamefont {Caspani}}, \bibinfo
  {author} {\bibfnamefont {J.}~\bibnamefont {Aza{\~n}a}}, \ and\ \bibinfo
  {author} {\bibfnamefont {R.}~\bibnamefont {Morandotti}},\ }\href
  {https://doi.org/10.1038/nature22986} {\bibfield  {journal} {\bibinfo
  {journal} {Nature}\ }\textbf {\bibinfo {volume} {546}},\ \bibinfo {pages}
  {622 EP } (\bibinfo {year} {2017})}\BibitemShut {NoStop}%
\bibitem [{\citenamefont {Imany}\ \emph {et~al.}(2018)\citenamefont {Imany},
  \citenamefont {Jaramillo-Villegas}, \citenamefont {Odele}, \citenamefont
  {Han}, \citenamefont {Leaird}, \citenamefont {Lukens}, \citenamefont
  {Lougovski}, \citenamefont {Qi},\ and\ \citenamefont {Weiner}}]{Imany:18}%
  \BibitemOpen
  \bibfield  {author} {\bibinfo {author} {\bibfnamefont {P.}~\bibnamefont
  {Imany}}, \bibinfo {author} {\bibfnamefont {J.~A.}\ \bibnamefont
  {Jaramillo-Villegas}}, \bibinfo {author} {\bibfnamefont {O.~D.}\ \bibnamefont
  {Odele}}, \bibinfo {author} {\bibfnamefont {K.}~\bibnamefont {Han}}, \bibinfo
  {author} {\bibfnamefont {D.~E.}\ \bibnamefont {Leaird}}, \bibinfo {author}
  {\bibfnamefont {J.~M.}\ \bibnamefont {Lukens}}, \bibinfo {author}
  {\bibfnamefont {P.}~\bibnamefont {Lougovski}}, \bibinfo {author}
  {\bibfnamefont {M.}~\bibnamefont {Qi}}, \ and\ \bibinfo {author}
  {\bibfnamefont {A.~M.}\ \bibnamefont {Weiner}},\ }\href {\doibase
  10.1364/OE.26.001825} {\bibfield  {journal} {\bibinfo  {journal} {Opt.
  Express}\ }\textbf {\bibinfo {volume} {26}},\ \bibinfo {pages} {1825}
  (\bibinfo {year} {2018})}\BibitemShut {NoStop}%
\bibitem [{sm()}]{sm}%
  \BibitemOpen
  \href@noop {} {}\bibinfo {note} {See Supplemental Material for
  discussion of the coherence and stability conditions, the reduced density
  matrices of the output state and of the W state, the measurement strategy,
  and the output state fidelity and purity, which includes Refs.~[37-39].}\BibitemShut {Stop}%
  
\bibitem{Parashar2009}P. Parashar and S. Rana, Phys. Rev. A \textbf{80}, 012319 (2009).

\bibitem{Lipinska2018}V. Lipinska, G. Murta, and S. Wehner, Phys. Rev. A \textbf{98}, 052320 (2018).

\bibitem{Chuang} M. A. Nielsen and I. Chuang, \textit{Quantum computation and quantum information}, Cambridge University Press (2000), page 100.
\end{thebibliography}
\end{document}